\documentclass[]{aa}
\usepackage{natbib}
\usepackage{psfig}
\def\HH{{\rm H}_2}
\def\nH2{{\rm n}({\rm H}_2)}
\def\NH2{{\rm N}({\rm H}_2)}
\def\pccc{{\rm cm}^{-3}} \def\pcc{{\rm cm}^{-2}}

\def\Tstar#1 {\mbox{${\rm T}_{\rm #1}^*$}}
\def\Tsub#1 {\mbox{$T_{\rm #1}$}}
\def\Tk  {\Tsub k }

\def\Texc {\Tsub exc }

\def\Tcmb{\Tsub cmb }

 \def\arcmin{\mbox{$^{\prime}$}}

\def\degr{$^{\rm o}$}
\def\p{$^+$}

\def\hcop{\mbox{{HCO\p}}} 
\def\chp{\mbox{CH\p}}
\def\cth{\mbox{C$_3$H}}
\def\cch{\mbox{C$_2$H}}
\def\cfh{\mbox{C$_4$H}} \def\CnHm{\mbox{C$_n$H$_m$}} 

\def\h13cop{\mbox{{H$^{13}$CO\p}}}
\def\nnhp{\mbox{N$_2$H\p}}
\def\c3h2{\mbox{C$_3$H$_2$}}
 
 \def\R0{R$_0$}

\def\ddeg{{}^\circ\kern-.1em}

\def\kms{\mbox{km\,s$^{-1}$}}

\def\E#1{\,10^{#1}}
\def\P#1,{$\nH2\Tk~=~#1\times~10^4~\pccc$~K}
\def\ec#1,#2,#3,{#1\,(#2)\E{#3}}

\def\zoph{$\zeta$ Oph}
\def\methCN{\mbox{CH$_3$CN}}
\def\H3{\mbox{H$_3$}}
\def\ammon{\mbox{N\H3} }
\title{Comparative Chemistry of Diffuse Clouds 
II:  CN, HCN, HNC, \methCN\ \& \nnhp}

\author{H. Liszt\inst{1}\ and R. Lucas\inst{2}}
\institute{National Radio Astronomy Observatory,
           520 Edgemont Road,
           Charlottesville, VA,
           USA 22903-2475
\and       Institut de Radioastronomie Millim\'etrique,
           300 Rue de la Piscine,
           F-38406 Saint Martin d'H\`eres,
           France}

\begin{document}
\date{Received 18 October 2000 /Accepted 26 January 2001}
\offprints{H. S. Liszt}
\mail{hliszt@nrao.edu}
\abstract{
Using the Plateau de Bure interferometer, we observed the $\lambda$3mm
absorption lines of CN, HCN and HNC from some of the diffuse clouds 
which lie toward our well-studied sample of compact extragalactic 
mm-wave continuum sources.  The column densities of these species all 
vary by a factor of about fifty and are prominent in only a limited 
subset of the clouds seen in the most ubiquitous species such as
OH,\hcop, \cch\ and \c3h2.  We searched unsuccessfully for \methCN\ 
and \nnhp, which are underabundant compared to dark clouds, by factors of 
at least 10 and 100, respectively.
The CN-HCN-HNC column densities vary strongly and non-linearly with 
N(\hcop), for example, which probably best represents their variation 
with $\HH$ as well. But their abundances are very tightly and linearly 
coupled to each other, varying in fixed proportion, as is the 
case for OH and \hcop, and (only slightly more loosely) for \cch\ and 
\c3h2.  Having measured one, it is hardly necessary to observe the two 
others in this group.  We find
$\langle$N(HNC)/N(HCN)$\rangle = 0.21\pm0.05$,
$\langle$N(CN)/N(HCN)$\rangle = 6.8\pm1$.
Such a small N(HNC)/N(HCN) ratio is typical of warmer gas in darker,
denser environments, and is consistent with the notion of molecular
formation in warmer media. The 6.8:1:0.21 ratio in diffuse gas is 
very different from TMC-1 (6.8:4.5:4.5) where HCN and HNC are relatively
much more abundant.
It seems likely that the sequence of features with increasing column 
densities of the CN-HCN family or \CnHm-family molecules in diffuse gas
actually represents  a  series of gas parcels of increasingly higher 
$\HH$-fraction, number density, and molecular abundances,  
occurring over a relatively narrow interval of total hydrogen column density.  
Our experiment seems to have caught many molecules in the act of turning 
on -- and turning each other on -- in the diffuse interstellar medium.
}
\maketitle

\section {Introduction.}

\begin{figure}
\psfig{figure=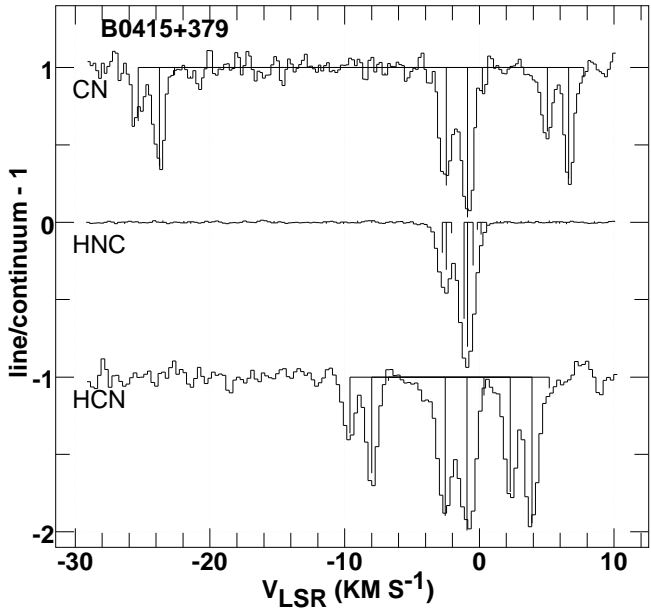,height=8cm}
\caption[]{CN, HCN and HNC absorption spectra toward B0415+379 (3C111) 
showing the basic hyperfine structure in a fit involving three kinematic
components (one of which is rather weak) for each species.}
\end{figure}

The presence of interstellar CN was shown by \cite{McK40} who identified
the R(0) line at $\lambda$387.4nm (see also \cite{SwiRos37} and \cite{Ada41}).
The R(1) line being undetectable, and taking as an upper limit on its 
equivalent width a value one-third that of R(0), he calculated the
``effective'' or ``rotational''  temperature of interstellar space to 
be no more than 2.7 K.  McKellar qualified his conclusion, inquiring 
``if, indeed, the concept of such a temperature in a region with so low
a density of both matter and radiation has any meaning.''  Of course it
does have meaning, and CN has long been used to derive the
temperature of the cosmic microwave background at a wavelength of 2.7mm,
the contribution to the rotational excitation from collisions often being
negligibly small 
\citep{Tha72,CraHeg+89,BlaVan91}.  
Both CH and \chp , the other early molecules, are too light for their 
rotational structure to have been used in this way.

Sixty years after the discovery of the CN lines we are still finding out
what their presence in diffuse clouds actually means.  Although it has not
generally been regarded as much of a challenge to come up with rather 
conventional chemical schemes which account for the observed abundances of CN  
\citep{FedStr+94,BlaVan86}, these models beg larger questions which are the 
subject of this series of papers.  Earlier we showed that the abundances
of the closely related species HCN and HNC are much larger than expected
in diffuse clouds seen toward compact extragalactic mm-wave continuum
sources at galactic latitudes of 2-15\degr\ \citep{LucLis93,LucLis94}.
Given that the proximate source of CN in conventional models is the
breakup of HCN and/or HNC and their immediate antecedents, it follows
that our understanding of CN is at best accidental 
(see also \cite{CraWil97}).

In the first paper in this new series \citep{LucLis00}, we discussed the 
comparative chemistry of species in the \CnHm-family of hydrocarbons whose 
most abundant known members are CH, \cch\ and \c3h2.  Here, we present results 
for the cyanogen- or CN-bearing molecules CN, HCN and HNC, based on a modest 
survey of $\lambda3$mm absorption from clouds occulting a subset of 
our wider sample of compact extragalactic continuum sources \citep{LucLis96}.  
We find a remarkably tight linear correlation among all three, such that
it is hardly necessary to observe the other two once the abundance of
any one has been found: a similarly tight linear relation exists between
OH and \hcop\ \citep{LisLuc94,LucLis96,LisLuc00} and a somewhat looser 
but still quite good one exists between \cch\ and \c3h2.  By contrast, 
intercomparisons among the different families (HCN or \cch\ $vs.$ \hcop;
HCN $vs.$ \cch) reveal a high degree of correlation, but sharply non-linear
behaviour.  In general, all species except OH increase in abundance
more rapidly than linearly with increasing N(\hcop) at
N(\hcop) $\ga 10^{12}~\pcc$ .  Analogous
non-linear behaviour in the optical regime is seen in plotting
N(CN) $vs.$ N(C$_2$) (power-law slope around 1.4, see below) or
either of these against N$(\HH)$, but tight linear relationships
between species are quite lacking in optical absorption studies.

The manner of taking and analyzing the present observations is discussed 
in Section 2.  The main results of this work are discussed in Section 3,
and a discussion of what is known of the chemistry of simple CN-bearing
molecules is presented in Section 4.

\begin{table}
\caption[]{Background sources and profile rms}
{
\begin{tabular}{lccccc}
\hline
Source&l& b & $\sigma_{l/c}$ &$\sigma_{l/c}$&$\sigma_{l/c}$ \\
      &\degr  &  \degr  & HCN & HNC & CN \\
\hline
B0212+735 & 128.93  & 11.96 & 0.040 & 0.066 &0.108\\
B0355+508 & 150.38  & $-$1.60 & 0.022 &0.006 &0.067\\
B0415+379 & 161.68 & $-$8.82 & 0.046&0.005&0.049\\
B0528+134 &  191.37 &$-$11.01 & 0.016&0.023&0.024\\
B1730$-$130&  12.03 & 10.81 & 0.009& 0.009 & 0.020\\
B2200+420 &  92.13 &$-$10.40 & 0.019& 0.052&0.069 \\
B2251+158&  86.11 &$-$38.18 & 0.008 & 0.014&\\
\hline
\end{tabular}}
\\
\end{table}

\section{Observations.}

\begin{figure}
\psfig{figure=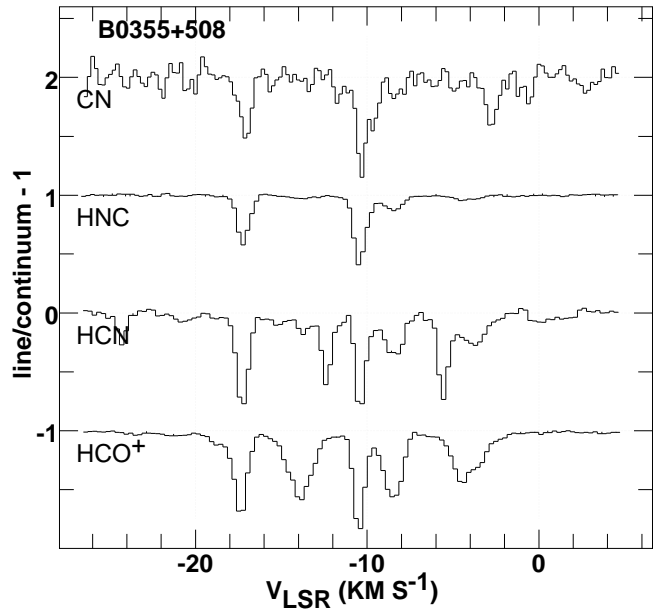,height=8cm}
\caption[]{\hcop, CN, HCN and HNC absorption spectra toward B0355+508. 
The HNC spectrum, in which the hyperfine structure only widens and does
not create (new) components, most readily shows which of the \hcop\ features 
are present in the CN family.}
\end{figure}

The mm-wave absorption data discussed here were taken at the Plateau
de Bure Interferomer during the years 1993 - 1997.  Table 1 shows 
the list of background sources observed, along with their galactic 
coordinates and the rms noise levels achieved in the line/continuum 
ratio, which is also the rms error in optical depth in the optically 
thin limit.  The mode of observation was identical to that described 
previously by \cite{LucLis00} and in our other prior work.  All data used
for the present survey were taken with a channel separation of 78 kHz 
and a channel width (FWHM) of  140 kHz, corresponding to velocity 
intervals which are given just below for the various individual species.

We have never succeeded in detecting emission from HCN, the most optically 
thick of the species discussed here, from the clouds occulting any of our 
extragalactic continuum sources. Thus we derive all column densities 
in the limit of no collisional excitation, i.e. \Texc\ = \Tcmb .  
Very weak HCN emission (0.02 K) was seen 
some 30\arcmin\ South of \zoph\ at the local \hcop\ emission peak 
(0.1 K; \cite{Lis97}) but CN emission is absent at very low levels 
toward the star itself 
{\citep{CraHeg+89,KopGer+96}}.

\subsection{CN}

We observed the (1/2-1/2), (3/2-5/2) and (1/2-3/2) hyperfine components of 
the N=0-1 absorption line near 113.490GHz, which have LTE line strengths in
the proportion 988:3333:1235.   As shown in Figure 1, the lines are split by 
3-11 MHz and so are separately resolved.  Assuming equilibrium with the 
cosmic microwave background, the CN column density is related to the 
integrated optical depth of the N=0-1 absorption as 
N(CN) $= 2.27 \times 10^{13} ~\pcc\int\tau_{0-1} dv$.

Results of Gaussian fitting are given in Table A1 
{(see Section 2.6).}
The channel separation and FWHM of the data are 0.206 \kms\ and 0.370 \kms, 
respectively.  

\subsection{HCN}


We observed the  F=1-1, F=2-1 and F=1-0 hyperfine components of the N=0-1
transition near 88.63GHz, with LTE line strengths in the ratio 1:5:3.  As
shown in Figure 1, the lines are split by 1-3 MHz and are separately
resolved, though often confused due to blending with other features.  Assuming
equilibrium with the cosmic microwave background, the HCN column density is
related to the optical depth of the N=0-1 absorption integrated over
all components as N(HCN) $ = 1.89 \times 10^{12} ~\pcc\int\tau_{0-1} dv$.
Results of Gaussian fitting are given in Table A2 
{(see Section 2.6). }
The channel separation and FWHM of the data are 0.264 \kms\ and 0.474 \kms,
respectively.

\subsection{HNC}


We observed the  F=1-1, F=2-1 and F=1-0 hyperfine components of the N=0-1
transition near 90.66GHz, with LTE linestrengths in the ratio 1:5:3.  As shown
in Figure 1, the hyperfine splittings of the J=0-1 absorption of
HNC are smaller than typical linewidths, and so are observed as a broadening
of the observed components rather than as separate kinematic features.  
Assuming equilibrium with the cosmic microwave background, the HNC column 
density is related to the integrated optical depth of the J=0-1 absorption as
N(HNC) $ = 1.76 \times 10^{12} ~\pcc\int\tau_{0-1} dv$.
Results of Gaussian fitting are given in Table A3
{(see Section 2.6).}
The channel separation and FWHM of the data are 0.256 \kms\ and 0.463 \kms,
respectively.

\subsection{\methCN}

We observed a band centered at 91.9871 GHz, $i.e.$ at the frequency of the 
strongest hyperfine component of the J=5-4 transition in the K=0 ladder,
toward B0415+379, B0528+134, B1730-130, and B2251+158.  This band sampled 
as well the same J-transition in the K=1 ladder, whose intensity should be 
smaller by a factor 0.076 in equilibrium with the cosmic microwave background;
yet higher K-ladder transitions falling in the band should have truly 
negligible population.  Although the spectrum is mildly suggestive of 
a detection toward 3C111, it is not statistically 
significant above the $2\sigma$ level.  The spectrum toward 3C111
had an rms noise in the line/continuum ratio of 0.004, leading to a 2-sigma 
upper limit on the integrated optical depth of $\int\tau dv < 0.014 $ \kms, or
N(\methCN) $< 2.4 \times 10^{11}~\pcc$, N(\methCN)/N(\hcop) $<$ 0.021.  
These are the results quoted in Table 3 and discussed below.  \methCN\ is 
underabundant compared to the composition of dark clouds like TMC-1 and L134N.

\subsection{\nnhp}

We observed a band around 93.1738 GHz, the location of the J=1-0 transitions
of \nnhp, toward B0355+508, B0415+379, B0528+134, B1730-130, and B2251+158,
but detected no absorption.  The best limit is for the gas
occulting B0415+379, and this is the 
{datum}
used in Table 3 and discussed 
below.  The spectrum toward 3C111 had an rms noise in the line/continuum 
ratio of 0.0025, leading to a 2-sigma upper limit on the integrated optical 
depth of $\int\tau dv < 0.016 $ \kms, or 
N(\nnhp) $< 2.6 \times 10^{10}~\pcc$, N(\nnhp)/N(\hcop) $< 0.0024$.  
By comparison, the other lines of sight did not furnish useful 
data except to confirm that we had not somehow missed large quantities of
\nnhp\ anywhere. \nnhp\ is much more strongly underabundant in diffuse
gas than \methCN.

\begin{figure}
\psfig{figure=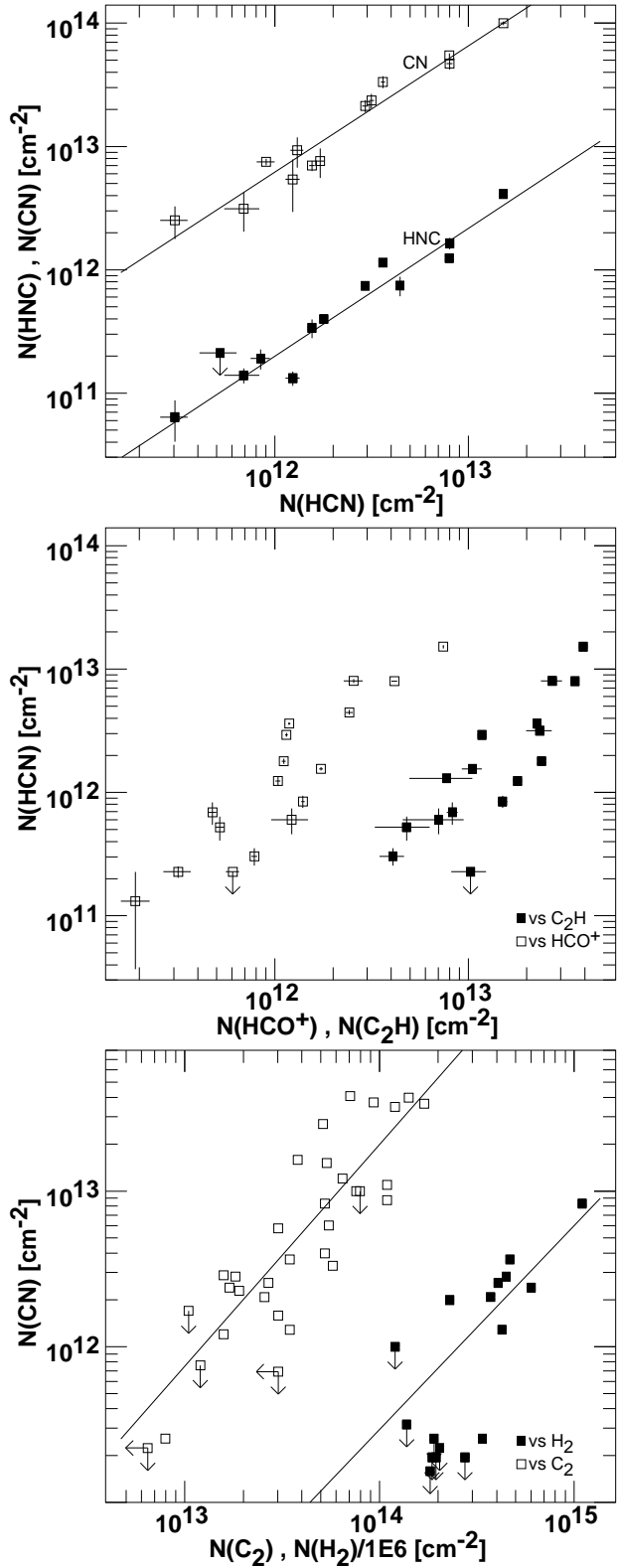,height=21cm}
\caption[]{HCN, CN, HNC, \hcop, \cch, C$_2$ and $\HH$ column densities 
compared at radio
and optical wavelengths.  Top:  N(CN) (open) and N(HNC) (filled) $vs.$ N(HCN).
Middle: N(HCN) $vs.$ N(\hcop) (open) and N(\cch) (filled).  Bottom, 
N(CN) $vs.$ N(C$_2$) and N($\HH$) at optical and $uv$-wavelengths, 
from the 1995 version of the compilation originally in \cite{FedStr+94}.}
\end{figure}

\subsection{Component fitting}

In Tables A1-3 of the Appendix we give the results of Gaussian decomposition, 
done simultaneously to whichever subset of the hyperfine structure was 
actually observed (all for HNC and HCN, and part for CN, see Figure 1). 
The optical depth quoted at line center pertains to the strongest
hyperfine component but was derived from a simultaneous fit to multiple 
kinematic and hyperfine components, the latter assumed to appear in the 
LTE ratio, as discussed in Section 2.4.  The integrated optical depth 
tabulated is the sum over all hyperfine components for a given kinematic 
feature, either derived directly (for HCN and HNC) or (for CN) by a simple 
scaling to account for that (smaller) fraction of the spectrum which was 
not actually observed.   

{We always fit the observed line profile (line/continuum ratio) using 
saturated gaussians of the form
$\exp(-\tau_0\times\exp(-(v-v_0)^2/2{\sigma_v}^2))$,
simultaneously considering all the accessible hyperfine structure.
Thus the optical depths and linewidths which are presented in the
Tables are consistent with each other and with the degree of 
saturation which occurs in the data.  The fitted linewidths
are not artifically increased by line saturation.}

As seen in Figure 2, the intrinsic kinematic structure along a given
line of sight can introduce substantial confusion as to which clouds
contain which molecule.  In each case, we used our \hcop\ profile
as a template and did a simultaneous fit, species-by-species, to all 
of the hyperfine and kinematic components which were observed.
The FWHM linewidths in Tables A1-A3 of the Appendix all include a 
contribution from the intrinsic broadening in the spectrometer
which, as noted above, has channel half-power widths somewhat wider 
than the channel separation itself.  Where the linewidths are plotted
(Figure 5), this contribution has been subtracted in quadrature 
from the tabulated value.  

\section{Systematics}

\subsection{Abundance and linewidth comparisons within the
family of small CN-bearing molecules}

Figure 3 at top shows that N(CN), N(HCN), and N(HNC) have the kind of
tight linear relationship previously demonstrated for OH and \hcop\
\citep{LisLuc96,LucLis96}.  The slopes of both regression lines in Figure 3
are indistinguishable from unity ($1.03\pm0.08$ in either case) with
mean ratios (weighted by the variance) of
$\langle$N(HNC)/N(HCN)$\rangle = 0.21\pm0.05$
and $\langle$N(CN)/N(HCN)$\rangle = 6.8\pm1.0$.

Figure 5 shows a comparison of the FWHM linewidths derived from
profile decomposition (see the Tables in the Appendix), corrected 
for finite resolution.  The HCN, HNC,
and \hcop\ linewidth intercomparisons show no deviation from the
condition of equal linewidths, but CN seems to have a slight tendency
to be narrower than HCN and a more marked tendency to be narrower than
\hcop.  CN is typically observed to be narrower than CH in optical
spectra, where the difference is usually interpreted as suggesting
a spatial segregration of CN into the denser core regions of a cloud
\citep{Cra95}.

In this work, CN was observed with a somewhat better spectral resolution, owing
to its higher rest frequency and the fixed nature of the channel widths
in the autocorrelator. Although it seems unlikely that resolution dominates the
linewidth comparison in Figure 5 -- HCN and \hcop\ are observed at nearly the
same frequency and resolution --  this should certainly be checked.
If CN really is narrower than HCN, this is extremely interesting because
the tight correlation of their column densities militates against any
really large-scale spatial segregation, such as isolating CN in a denser core.
Instead, there would have to be a mix of species and conditions 
throughout the gas such that line of sight average column densities vary 
in fixed proportion, and the CN linewidth would be smaller throughout
the cloud, or at least down to scales much finer than we can discern
directly.

\subsection{Abundance comparisons across families of species observed 
at mm-wavelengths}

Figure 3 in the middle shows the variation of HCN with respect to
\hcop\ and \cch.  As we have shown before, comparisons across chemical
families reveal sharply non-linear behaviour; although the variations
of HCN with \hcop\ and \cch\ may appear similar, the \cch-\hcop\ relationship
is also non-linear, with the same increase in N(\cch) at N$(\hcop) \approx
1 \times 10^{12}~\pcc$ (Paper I).  \cch\ however has something
of a plateau at small N(\hcop), where the mean of N(\cch)/N(\hcop) is high.
The N(HCN)/N(\cch) ratio varies from (roughly) 0.08 to 0.3, increasing 
with column density.

Taken as unweighted averages over all observed features, we find
$<$N(HCN)/N(\hcop)$> = 1.47\pm0.86$, $<$N(HCN)/N(\cch)$> = 0.155\pm0.094$;
in Paper I we quoted  $<$N(\cch)/N(\hcop)$> = 14.5\pm6.7$ for the unweighted
mean.  When the averages are weighted by the fractional errors of the column
density, favoring higher column density sources, we have 
$<$N(HCN)/N(\hcop)$> = 1.97\pm0.43$ and 
$<$N(HCN)/N(\cch)$> = 0.23\pm0.57$.

\begin{figure}
\psfig{figure=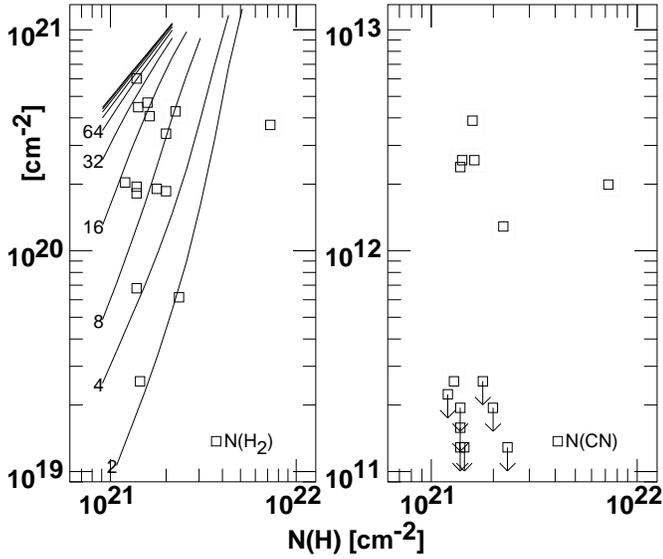,height=7.3cm}
\caption[]{N($\HH$) (left) and N(CN) plotted against the total
H-column density N(H) = N(H I) + 2*N($\HH$) for lines of sight with N(CN)
measured at optical wavelengths (as in the bottom panel of Figure 3).
For convenience, datapoints corresponding to sightlines with 
optically-measured CN column densities have been outlined.  Superposed 
in the left-hand panel are theoretical $\HH$-formation curves for
uniformly-illuminated gas clots of constant total density n(H) 
as given ($\pccc$) at their left-hand edges.  N(H) and N($\HH$) are 
from \cite{SavDra+77}.}
\end{figure}

\subsection{Insights from optical absorption data}

Figure 3 at bottom shows optically-determined CN, C$_2$, and 
$\HH$ column densities from the updated version of the summary
compilation of \cite{FedStr+94}.  It is lamentable that there are so few
lines of sight for which both molecular hydrogen and CN have been observed.
In any case, although the relationships in Figure 3 at the bottom are
non-linear (power-law slopes of 1.42 and 1.31 reading from left to right)
they provide a useful calibration of absolute abundances.  For instance,
at N(CN) = $2.1\times10^{12}~\pcc$, the fitted curve gives N($\HH$)
= $4.2\times10^{20}~\pcc$.  This can be transferred to the fits at the
top of Figure 3 to infer N(HCN) = $3.1\times10^{11}~\pcc$ and then to
Figure 3 at the middle for (very) approximately
N(\hcop) $\approx 7.2\times10^{11}~\pcc$, or
N(\hcop)/N($\HH$) $\approx 1.7\times10^{-9}$. This is near the
single value N(\hcop)/N($\HH$) =  $2\times10^{-9}$ that we have used
for a variety of purposes recently \citep{LisLuc00}, most especially to 
show that the observed amounts of CO in diffuse clouds follow directly 
from recombination of this amount of \hcop\ with electrons in standard 
cloud models.

As inferred from this train of thought, the \hcop\ abundance is found to
be slightly smaller ($\approx 1.5\times10^{-9}$) at
N($\HH) = 1.0\times10^{21}~\pcc$. But from this comparison we gain the
new insight that the abrupt increases in the abundances of so many species
at N$(\hcop) \ga 10^{12}~\pcc$ are actually 
{occurring}
at N($\HH) \approx 0.3-0.5\times10^{21}~\pcc$.

\subsection{Linear and non-linear functional dependencies}

It is important to understand which parameters and physical conditions
are varying in Figure 3.  For instance, at the top there is a factor 50 
variation in N(CN), N(HCN) and N(HNC), but much of this -- perhaps a
factor 20-30  -- occurs over a much smaller range in N(\hcop):   
a factor 5-6.  Thus the total column density of gas probably changes
far less than might otherwise be guessed. At the bottom, N(CN) seems to 
vary by a factor ten or so at N($\HH$) $\approx 3\times10^{20}~\pcc$, 
and less rapidly, but still faster than linearly, at higher N($\HH$).  

To tie everything together somewhat more tightly, we show in Figure 4 
the variation of N(CN) and N($\HH$) with total gas column density 
N(H) = N(H I) + 2N($\HH$) \citep{SavDra+77} for those sightlines 
in Figure 3 where CN has been measured at optical wavelengths.  
The point is that, while N(CN) and N($\HH$) are strongly coupled at
higher N($\HH$) in Figure 3, neither N(CN) nor N($\HH$) increases
with N(H) or extinction along those same directions.  Both N(CN) and N($\HH$) 
vary substantially over a very small range of N(H).
\footnote{In general the effect of extraneous atomic hydrogen is a rightward 
displacement in Figure 4, causing the data to appear consistent with 
$\HH$-formation at too low a local number density.  The anomalous point 
at large N(H) in Figure 4 for the sightline to $\rho$ Oph A is an 
extreme example.  No other sightline in the {\it Copernicus} sample 
had such a large N(H), for any value of N($\HH$).}

The theoretical $\HH$-formation curves superposed in Figure 4 
at left were constructed by sampling along the mean line of sight through 
a uniformly-illuminated spherical gas parcel of constant total density n(H)
\citep{LisLuc00}.
They suggest that those sightlines with larger N($\HH$) -- and so with 
larger N(CN) -- sample denser and perhaps darker gas with 
a higher local fraction of molecular hydrogen. This in turn leads to the
idea that there is some cascade of effects which accounts for the very 
strong changes in column density among the trace species, largely driven 
by the final approach of the molecular hydrogen fraction toward unity,
but perhaps having as its cause the change in some other agent. 

In the most classical terms, an increase in local number density would 
drive the molecular hydrogen fraction higher at relatively constant 
extinction, increasing the rate of formation of ancestor molecular 
ions like $\HH$CN$^+$, which would then recombine faster at higher 
density and (if the thermal pressure is even approximately conserved, 
as is almost the case in our models) somewhat lower temperature.   
It helps considerably that both the electron fraction and the abundance of 
the basic building material (C$^+$) are dominated by the 
C$^+$/C$^0$ ratio, which remains high in diffuse gas at the 
moderate density and extinction relevant here.  Earlier we showed
that even a constant abundance (relative to $\HH$) of an ancestor
ion could give rise to faster-than-linear increases in the abundance
of a product molecule {\it via} electron-ion recombination in a gas of
constant density, specifically N(CO) varying as N$(\HH)^2$ when 
arising from gas with constant N(\hcop)/N$(\HH$) \citep{LisLuc00}.

A related matter of interest is the extent to which the molecules we  
sample may exist in environments which are still partly atomic.  In Figure 
6 we show the variation of the $\HH$-fraction with position inside spherical 
gas parcels of density n(H) = 16 and 128 H-nuclei $\pccc$, for five values 
of the column density across the central line of sight through the parcel.  
At the panel to the right in this Figure, where the number density is large 
enough that weak \hcop\ emission 
might be seen, even low-extinction gas is mostly molecular: the lowest column 
density considered here is about that of a standard ``Spitzer'' HI cloud.  
But at low density, and when small molecular column densities are sampled, 
the ambient environment of hydrogen gas is likely to be largely atomic.

\begin{table}
\caption[]{Relative abundances 10$^8 \times$ N()/N($\HH$)}
{
\begin{tabular}{lcccc}
\hline
Species&\zoph& This work& TMC-1 & BD-G \\
\hline
OH&10 &7&30  &10 \\
CO&480&&8000  &41 \\
\hcop&0.2&0.2&0.8&0.009 \\
C\p&26100&&& 89100 \\
C&700&&&720 \\
C$_2$&3.3&&&3.7 \\
C$_3$&$<0.012$&&&$10^{-5}$ \\
CH&5.4&1-2&2&3.9 \\
\chp&6.3&&&0.006 \\
\cch&&1.3-4&7&0.4 \\
\cth&&$<0.02$&0.05& \\
\c3h2&&0.05-0.15&3& \\
\cfh&    &&2.0&\\
NH  &0.19&&&0.10 \\
CN&0.54&1.0-3.5&3&0.30\\
HCN&(0.079)&0.14-0.5 &2&0.007\\
HNC    &(0.016)&0.03-0.10 &2   &\\
\methCN&       &$<0.004$   &0.1&\\
\nnhp  &       &$<0.0004$  &0.2& \\
\hline
\end{tabular}}
\\
``This work'' represents this work,  results from Paper I 
( for \CnHm-family species)), unpublished CH data and 
a scaling N(OH) = 35 N(\hcop) \\
Results for TMC-1 are from \cite{OhiIrv+92}  \\
BD-G is \zoph\ model G of \cite{BlaVan86}\\
For \zoph, C\p\ is from \cite{CarMat+93}, NH is from 
\cite{CraWil97} and HCN and HNC are normalized to CN \\
\end{table}

\begin{figure}
\psfig{figure=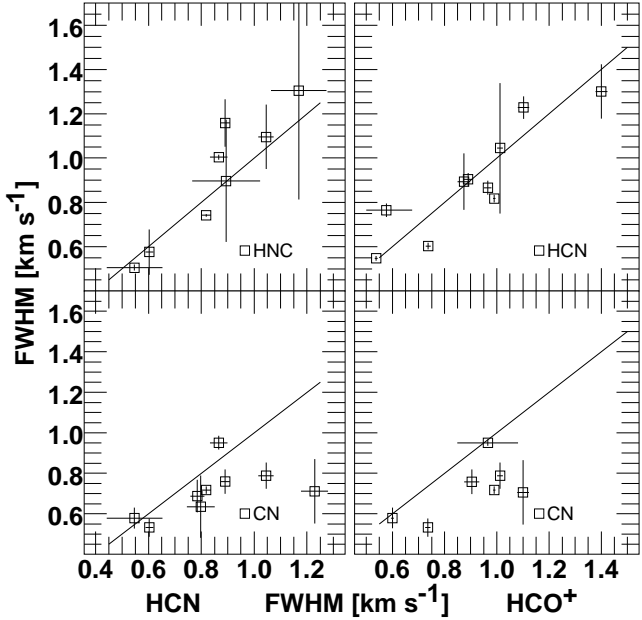,height=8.2cm}
\caption[]{Comparison of FWHM linewidths.  Left: HNC (top) and
CN (bottom) $vs.$ HCN.  Right:  HCN (top) and CN (bottom) $vs.$ \hcop.
HCN, HNC, and \hcop\ have statistically indistinguishable widths.  CN
is noticeably narrower than \hcop .}
\end{figure}

\section{Chemistry of simple nitrogen-bearing molecules}

\subsection{CN, HCN and HNC}

Shown in Table 2 is a comparison across diffuse and dark
gas of the chemical abundances of the hydrocarbons studied in Paper I and
species relevant to the present work (NH, CN, HCN, HNC).  The tabulation
indicates that the abundances of HCN and HNC increase very substantially
in moving from diffuse to dark gas while CN increases in abundance only 
slightly.

As represented in the last column of Table 2 by the work of \cite{BlaVan86}, 
models of moderate-density quiescent gas-phase diffuse cloud chemistry 
provide approximately the correct amounts of NH and CN, but fail to 
provide the 
inferred amount of HCN (which we derive by scaling N(CN)) by a factor of 10.   
However even this degree of agreement may be illusory.  The  \cite{BlaVan86} 
results were computed for a very much larger carbon abundance than is now 
accepted (see the entry under \zoph\ from \cite{CarMat+93}) and the claim 
is made (see the discussion of \cite{CraWil97}) that ordinary gas phase 
diffuse cloud chemistry fails by a factor of about 40 to reproduce the 
observed N(NH) and N(CN). If so, the predicted amounts of most molecules 
would be much smaller than are indicated in the last column of Table 2.

It seems generally accepted that formation of the CN-bearing molecules
proceeds $via$ the reaction C$^+ + {\rm NH} \rightarrow {\rm CN}^+ + $H, 
followed by a series of hydrogen abstractions beginning with
CN$^+ + \HH \rightarrow {\rm HCN}^+ +$ H, and terminating (forming one
of the observed species CN, HCN, HNC) after one of many possible electron-ion
recombinations such as $\HH$CN$^+ + {\rm e} \rightarrow {\rm HCN} + $H 
\citep{BlaVan86,FedStr+94,CraWil97}. However, the particulars are very much open
to question since the needed gas-phase reactions do not proceed nearly 
fast enough under typical quiescent conditions in diffuse clouds.

Quiescent gas phase chemical models make NH by cosmic ray ionization of N,
followed by hydrogen abstraction in the reaction of N\p\ and $\HH$. This 
and all
other gas phase NH formation reactions are discounted by \cite{CraWil97} 
in favor of  formation on grains; they also argue that the linewidth of 
NH, which is  typical, rules out any formation mechanism involving 
translationally hot particles in the gas phase, like ions (N\p) accelerated 
in MHD shocks.

\begin{figure}
\psfig{figure=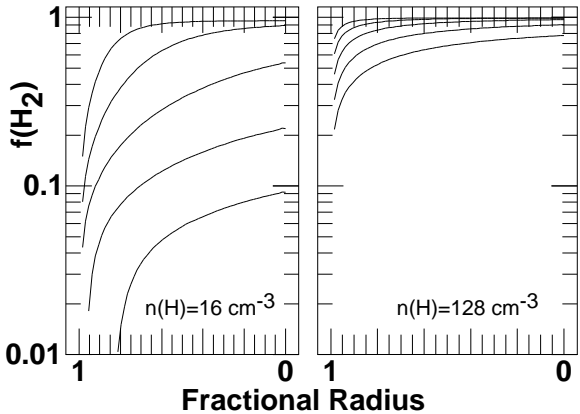,height=5.3cm}
\caption[]{Theoretical $\HH$-formation curves for uniformly-illuminated
spherical gas 
parcels of constant density n(H) = n(HI)+2n($\HH$) = 16 $\pccc$
and 128 $\pccc$. In each panel the fraction of nuclei in 
$\HH$, f($\HH$) = 2 n($\HH$)/n(H), is plotted as a function of 
fractional radius for five values of the total column density
along the line of sight intersecting the center of the parcel, 
N(H) = 0.25, 0.5, 1, 2 and 4 $\times10^{21}~\pcc$.}

\end{figure}

Whether the NH may be formed in gas-phase reactions with fast-moving 
particles is of interest because a similar formation route is cited 
as the likely source of the unexplained high abundances of \hcop\ 
($i.e.$ C\p\ + OH $\rightarrow$ CO\p\ + H) and CH\p\ ($via$ C\p\ + $\HH$)
\citep{FalPin+95,HogDeG+95,FedRaw+96,JouFal+98}.
One of these pathways also seems to be needed to explain the observed 
amounts of CO, as well,  and they would help greatly in alleviating 
the apparent shortfall of the CN-bearing species (through the reaction 
of C\p\ and NH). If such processes do not contribute to alleviating the 
apparent shortfall of NH, it seems possible that they might not be needed 
at all to form the CN-bearing species.  Alternately, if they do function, 
they would represent strong sources and sinks of NH not considered by 
\cite{CraWil97}.

Whether the linewidth of NH rules out an enhanced gas phase formation
rate really awaits a calculation of reaction dynamics and subsequent
molecular thermalization in the ambient medium \citep{BucGli99}.  For 
molecules which are formed hot -- which includes both those which are 
formed from translationally hot reactants and those which carry away 
significant amounts of energy from 
exoergic chemical reactions -- only those which react quickly will 
disappear before reaching kinetic equilibrium and so exhibit a broader
linewidth.  CH\p, for instance, reacts rapidly with both $\HH$ and
electrons, while NH does not.

\subsection{\methCN}

Table 2 shows that \methCN\ is underabundant in diffuse gas
relative to TMC-1, by at least a factor of twenty, or somewhat
less, 6, if its abundance relative to \hcop\ is considered.  Given
that easily-observed species like \c3h2\ and HCN are underabundant
by a similar factor, the limit on \methCN\ is not that strong
and should probably not be considered too surprising.  Both of the
chemical sequences leading to \methCN\ in dark gas \citep{NejWag99},
(C, C$\HH$, HCN, H$_4$C$_2$N$^+$, \methCN) at early times or
(C$_3$H, C$_3$N, HC$_3$N$^+$, HCN,  H$_4$C$_2$N$^+$, \methCN)
later on, run through HCN.

\subsection{\nnhp}

\nnhp\ is one of several molecules (\nnhp, \ammon, SO$_2$) which are 
considered to form only very late in the chemical evolution of
dark clouds \citep{NejWag99}.  Observationally, \ammon\ at least
is much more widespread than \nnhp\ and has been seen toward B0355+508
by \cite{Nas90} and toward 3C111 in our own unpublished work.  However, 
unpublished VLA results communicated by A. Marscher indicate that 
reliable \ammon\ abundances await a survey with the upgraded
K-band receivers at the VLA, and that neither Nash's profiles nor ours
(all taken with single dishes), are trustworthy.  Unfortunately,
\ammon\ column densities are also unavailable from Marscher's VLA
profiles, because only the (1,1) lines were observed.

In molecular clouds, \nnhp\ forms in appreciable quantities at late times
$via$ \H3\p $+$ N$_2$ $\rightarrow$ \nnhp\ $+$ $\HH$, {\it i.e.} when the 
ionization fraction has dropped and when CO has been depleted from the 
gas phase, encouraging a higher abundance of \H3\p. It is known for being 
detectable only in darker regions.  

Toward 3C111, N(\nnhp)/N(\hcop) $< 0.002$ (2-sigma) as compared to 0.25 in 
TMC-1.  This is by far the best limit we have ever been able to set on an 
undetected species.  
{3C111 is actually seen through a small hole (region
of lower than average extinction) in an outlying cloud in the Taurus
cloud complex; a discussion of the viewing geometry is given by 
\cite{LucLis98}.}

\section{Summary}

We showed that the abundances of CN, HCN and HNC vary in fixed proportion
6.8:1:0.21 respectively.  This is rather different from conditions 
in TMC-1, where their abundances are in the ratio 6.8:4.5:4.5.  
The fractional abundances of HCN and HNC increase substantially between 
diffuse and dark gas but that of CN increases only by factor of a few.
Nonetheless, HCN and HNC are much more abundant in diffuse gas than 
predicted by models of quiescent low-density gas-phase chemistry,
with an intrinsic variation over the approximate
range N(HCN)/N($\HH) \approx 1.4 - 5.0 \times\ 10^{-9}$ for 
 N(\hcop)/N($\HH) = 2.0 \times\ 10^{-9}$.  The latter value is 
confirmed by comparing our measurements of CN, HCN, and \hcop\ with 
the lamentably slight body of optical absorption data for lines of sight
where both CN and $\HH$ have been observed.  

The linewidths of HCN, HNC, and \hcop\ are sensibly identical, but CN
shows a slight tendency to be narrower than HCN and a more pronounced
tendency to be narrower than \hcop.  If the CN and HCN are strongly
coupled in column density but different kinematically, this argues against
a spatial segregation of CN into denser regions of a cloud (core), such
as is usually invoked to explain the narrower width of CN absorption
lines seen in the optical.  Instead, clouds would have to be a mixture
of kinematically (and chemically) distinct regions all occurring many 
times along any one line of sight.
 
In comparing the abundances of species in the CN family with the hydrocarbons
discussed in Paper I (principally \cch\ and \c3h2) or with \hcop, we find
strong correlations but sharply non-linear variation.  Both the CN-species 
and hydrocarbons  show a strong increase in column density around 
N(\hcop) $= 10^{12}~\pcc$ and all comparisons
across families show relative abundance changes by factors of 3 or more.
We note that \cch\ and \c3h2\ vary in fixed proportion with each other,
though not as tightly as for \hcop\ and OH, or as seen here.

For those sightlines where the column densities of CN, HI, and 
$\HH$ have been measured optically,  it appears that N(CN) 
turns on abruptly at N($\HH$) $ \approx 3\times10^{20}~\pcc$ and 
increases faster than linearly at larger N($\HH$), in a regime where 
the abundance of $\HH$ itself varies widely over a very small range 
in N(H) (or extinction).  Models of $\HH$-formation suggest that the 
hydrogen is still largely atomic in the gas sampled on sightlines with 
the smallest column densities of the trace species like CN.

Earlier \citep{LisLuc00}, we showed that 
N(CO) $ \propto N(\HH)^2 $ (approximately) when 
CO forms in a gas of constant density {\it via} recombination of \hcop,
and the latter is held in fixed proportion ($2\times10^{-9}$) to $\HH$.  
It follows that much stronger variations in trace abundances would occur 
if the higher molecular fractions along some sightlines -- reflected in higher 
N($\HH$) and N(\hcop) -- arise because the density is higher locally; the 
sequence of molecular abundances revealed in our work must at least partly 
reflect the tendency for higher-density regions to harbor higher 
$\HH$-fractions at a given total hydrogen column density. Of course this 
cannot be the whole story because the high abundances of the parent ions 
needed to reproduce our observations of their descendants cannot presently 
be explained by any agreed-upon physical scenario.  Equally important is 
to understand why the abundances of some species relative to $\HH$ vary 
so strongly ($e.g.$ \cch, \c3h2, CN, HCN, and HCN) while those of 
CH, OH, and \hcop\ seem nearly fixed.

We failed to find \methCN\ and \nnhp, meaning that they are at least 10 and
100 times less abundant (relative to \hcop) in diffuse clouds than in
TMC-1.  The limit on \methCN\ may not be very significant, given that
\c3h2\ and HCN are underabundant by about the same amount but are readily
seen along many lines of sight and that, in dark gas, the formation paths 
to \methCN\ pass through HCN at both early and late times.  The limit for 
\nnhp\ may be more interesting but this species is notoriously hard to 
detect even in translucent clouds of much higher density than we study.

We discussed what is known of the chemistry of the CN-HCN-HCN 
species.  Although there is a consensus that they form through some
chemical pathway involving NH, most likely interacting with C\p, the 
origin of NH and the mode of interaction are entirely uncertain.  

The next papers in this series will discuss the sulfur-bearing species 
CS, SO, SO$_2$, $\HH$S and HCS\p, and probably N\H3\ and $\HH$CO.

\begin{acknowledgements}

We thank Dr. John Black for a close reading of the manuscript.

The National Radio Astronomy Observatory is operated by AUI, Inc. under a
cooperative agreement with the US National Science Foundation.  IRAM is
operated by CNRS (France), the MPG (Germany) and the IGN (Spain). We owe the
staff at IRAM (Grenoble) and the Plateau de Bure our thanks for their 
assistance in taking the data.

\end{acknowledgements}

\bibliographystyle{apj}
\bibliography{ms10370b2,ms10370b1}

\begin{thebibliography}{30}
\expandafter\ifx\csname natexlab\endcsname\relax\def\natexlab#1{#1}\fi

\bibitem[{{Adams}(1941)}]{Ada41}
{Adams}, W.~S. 1941, ApJ, 93, 11

\bibitem[{{Black} \& {van Dishoeck}(1991)}]{BlaVan91}
{Black}, J.~H. \& {van Dishoeck}, E.~F. 1991, ApJ, 369, L9

\bibitem[{{Bucher} \& {Glinski}(1999)}]{BucGli99}
{Bucher}, M.~E. \& {Glinski}, R.~J. 1999, Mon. Not. R. Astron. Soc., 308, 29

\bibitem[{{Cardelli} {et~al.}(1993){Cardelli}, {Mathis}, {Ebbets}, \&
  {Savage}}]{CarMat+93}
{Cardelli}, J.~A., {Mathis}, J.~S., {Ebbets}, D.~C., \& {Savage}, B.~D. 1993,
  ApJ, 402, L17

\bibitem[{{Crane} {et~al.}(1989){Crane}, {Hegyi}, {Kutner}, \&
  {Mandolesi}}]{CraHeg+89}
{Crane}, P., {Hegyi}, D.~J., {Kutner}, M.~L., \& {Mandolesi}, N. 1989, ApJ,
  346, 136

\bibitem[{{Crawford}(1995)}]{Cra95}
{Crawford}, I.~A. 1995, Mon. Not. R. Astron. Soc., 277, 458

\bibitem[{{Crawford} \& {Williams}(1997)}]{CraWil97}
{Crawford}, I.~A. \& {Williams}, D.~A. 1997, Mon. Not. R. Astron. Soc., 291,
  L53

\bibitem[{{Falgarone} {et~al.}(1995){Falgarone}, {Pineau Des For\^ets}, \&
  {Roueff}}]{FalPin+95}
{Falgarone}, E., {Pineau Des For\^ets}, G., \& {Roueff}, E. 1995, A\&A, 300,
  870

\bibitem[{{Federman} {et~al.}(1996){Federman}, {Rawlings}, {Taylor}, \&
  {Williams}}]{FedRaw+96}
{Federman}, S.~R., {Rawlings}, J. M.~C., {Taylor}, S.~D., \& {Williams}, D.~A.
  1996, Mon. Not. R. Astron. Soc., 279, L41

\bibitem[{{Federman} {et~al.}(1994){Federman}, {Strom}, {Lambert}, {Cardelli},
  {Smith}, \& {Joseph}}]{FedStr+94}
{Federman}, S.~R., {Strom}, C.~J., {Lambert}, D.~L., {Cardelli}, J.~A.,
  {Smith}, V.~V., \& {Joseph}, C.~L. 1994, ApJ, 424, 772

\bibitem[{{Hogerheijde} {et~al.}(1995){Hogerheijde}, {De Geus}, {Spaans}, {Van
  Langevelde}, \& {Van Dishoeck}}]{HogDeG+95}
{Hogerheijde}, M.~R., {De Geus}, E.~J., {Spaans}, M., {Van Langevelde}, H.~J.,
  \& {Van Dishoeck}, E.~F. 1995, ApJ, 441, L93

\bibitem[{{Joulain} {et~al.}(1998){Joulain}, {Falgarone}, {Des Forets}, \&
  {Flower}}]{JouFal+98}
{Joulain}, K., {Falgarone}, E., {Des Forets}, G.~P., \& {Flower}, D. 1998,
  A\&A, 340, 241

\bibitem[{{Kopp} {et~al.}(1996){Kopp}, {Gerin}, {Roueff}, \& {Le
  Bourlot}}]{KopGer+96}
{Kopp}, M., {Gerin}, M., {Roueff}, E., \& {Le Bourlot}, J. 1996, A\&A, 305, 558

\bibitem[{{Liszt}(1997)}]{Lis97}
{Liszt}, H.~S. 1997, A\&A, 322, 962

\bibitem[{{Liszt} \& {Lucas}(1994)}]{LisLuc94}
{Liszt}, H.~S. \& {Lucas}, R. 1994, ApJ, 431, L131

\bibitem[{{Liszt} \& {Lucas}(1996)}]{LisLuc96}
---. 1996, A\&A, 314, 917

\bibitem[{{Liszt} \& {Lucas}(2000)}]{LisLuc00}
---. 2000, A\&A, 355, 333

\bibitem[{{Lucas} \& {Liszt}(1998)}]{LucLis98}
{Lucas}, R. \& {Liszt}, H. 1998, A\&A, 337, 246

\bibitem[{{Lucas} \& {Liszt}(1993)}]{LucLis93}
{Lucas}, R. \& {Liszt}, H.~S. 1993, A\&A, 276, L33

\bibitem[{{Lucas} \& {Liszt}(1994)}]{LucLis94}
---. 1994, A\&A, 282, L5

\bibitem[{{Lucas} \& {Liszt}(1996)}]{LucLis96}
---. 1996, A\&A, 307, 237

\bibitem[{{Lucas} \& {Liszt}(2000)}]{LucLis00}
---. 2000, A\&A, 355, 327

\bibitem[{{McKellar}(1940)}]{McK40}
{McKellar}, A. 1940, Publ. Astron. Soc. Pac., 52, 187

\bibitem[{{Nash}(1990)}]{Nas90}
{Nash}, A.~G. 1990, Astrophys. J., Suppl. Ser., 72, 303

\bibitem[{{Nejad} \& {Wagenblast}(1999)}]{NejWag99}
{Nejad}, L. A.~M. \& {Wagenblast}, R. 1999, A\&A, 350, 204

\bibitem[{{Ohishi} {et~al.}(1992){Ohishi}, {Irvine}, \& {Kaifu}}]{OhiIrv+92}
{Ohishi}, M., {Irvine}, W., \& {Kaifu}, N. 1992, in Astrochemistry of cosmic
  phenomena: proceedings of the 150th Symposium of the International
  Astronomical Union, held at Campos do Jordao, Sao Paulo, Brazil, August 5-9,
  1991. Dordrecht: Kluwer 1992, ed. P.~D. {Singh}, 171--172

\bibitem[{{Savage} {et~al.}(1977){Savage}, {Drake}, {Budich}, \&
  {Bohlin}}]{SavDra+77}
{Savage}, B.~D., {Drake}, J.~F., {Budich}, W., \& {Bohlin}, R.~C. 1977, ApJ,
  216, 291

\bibitem[{{Swings} \& {Rosenfeld}(1937)}]{SwiRos37}
{Swings}, P. \& {Rosenfeld}, L. 1937, ApJ, 86, 483

\bibitem[{{Thaddeus}(1972)}]{Tha72}
{Thaddeus}, P. 1972, Ann. Rev. Astrophys. Astron., 10, 305

\bibitem[{{Van Dishoeck} \& {Black}(1986)}]{BlaVan86}
{Van Dishoeck}, E.~F. \& {Black}, J.~H. 1986, Astrophys. J., Suppl. Ser., 62,
  109

\end{thebibliography}

\begin{appendix}

\section{Products of Gaussian fitting}

\begin{table*}
\caption[]{CN Absorption line decomposition products}
{
\begin{tabular}{lcccc}
\hline
Source & v    & $\tau_0$ & FWHM & $\int\tau dv$ \\
       & \kms &          &\kms  & \kms \\
\hline
B0212+735& 3.612(0.033)& 1.536(0.104)& 0.759(0.061)& 2.067(0.217) \\
B0355+508& -17.076(0.024)& 0.995(0.073)& 0.532(0.045)& 0.939(0.106) \\
         & -11.197(0.380)& 0.115(0.031)& 1.621(0.696)& 0.331(0.172) \\
         & -10.295(0.023)& 1.432(0.096)& 0.579(0.051)& 1.472(0.163) \\
         & -8.292(0.073)& 0.267(0.041)& 0.706(0.159)& 0.335(0.092) \\
B0415+379& -2.458(0.017)& 1.431(0.030)& 0.951(0.034)& 2.415(0.100) \\
         & -0.872(0.011)& 3.459(0.060)& 0.717(0.019)& 4.399(0.140) \\
         & 0.246(0.063)& 0.169(0.027)& 0.459(0.142)& 0.138(0.048) \\
B0528+134& 9.582(0.030)& 0.220(0.010)& 0.788(0.065)& 0.308(0.029) \\
B1730-130& 4.980(0.139)& 0.049(0.008)& 1.268(0.312)& 0.111(0.033) \\
2200+420& -0.974(0.038)& 0.852(0.056)& 0.687(0.080)& 1.038(0.139) \\
           & 0.019(0.069)& 0.367(0.046)& 0.635(0.154)& 0.413(0.113) \\
\hline

\end{tabular}}
\end{table*}

\begin{table*}
\caption[]{HCN Absorption line decomposition products}
{
\begin{tabular}{lcccc}
\hline
Source & v    & $\tau_0$ & FWHM & $\int\tau dv$ \\
       & \kms &          &\kms  & \kms \\
\hline
B0212+735& -0.092(0.044)& 0.234(0.031)& 0.578(0.099)& 0.259(0.056) \\
         & 3.481(0.011)& 2.584(0.037)& 0.890(0.018)& 4.406(0.107) \\
B0212+735& -0.104(0.047)& 0.238(0.031)& 0.605(0.104)& 0.276(0.060) \\
         & 2.655(0.101)& 0.204(0.032)& 0.811(0.143)& 0.318(0.075) \\
         & 3.517(0.019)& 2.750(0.044)& 0.805(0.026)& 4.241(0.154) \\
B0355+508& -17.205(0.009)& 1.345(0.036)& 0.603(0.015)& 1.555(0.057) \\
         & -13.770(0.049)& 0.191(0.010)& 1.301(0.123)& 0.477(0.051) \\
         & -10.406(0.006)& 1.828(0.032)& 0.547(0.010)& 1.915(0.049) \\
         & -8.397(0.027)& 0.386(0.011)& 1.228(0.051)& 0.907(0.046) \\
         & -4.632(0.073)& 0.163(0.007)& 2.104(0.154)& 0.655(0.056) \\
B0415+379& -3.975(0.274)& 0.107(0.016)& 1.867(0.486)& 0.382(0.115) \\
         & -2.518(0.016)& 2.540(0.045)& 0.866(0.032)& 4.216(0.174) \\
         & -0.926(0.011)& 5.442(0.074)& 0.818(0.017)& 8.532(0.212) \\
B0528+134& 9.609(0.014)& 0.411(0.007)& 1.045(0.029)& 0.824(0.027) \\
B1730-130& 4.914(0.070)& 0.094(0.006)& 0.894(0.128)& 0.161(0.026) \\
         & 5.983(0.199)& 0.034(0.005)& 0.986(0.382)& 0.065(0.027) \\
B2200+420& -1.426(0.018)& 1.111(0.015)& 0.785(0.027)& 1.670(0.063) \\
         & -0.527(0.035)& 0.451(0.014)& 0.799(0.053)& 0.690(0.050) \\
B2251+158& -9.545(0.048)& 0.054(0.003)& 1.169(0.105)& 0.120(0.013) \\
B0355+508& -17.207(0.007)& 1.390(0.017)& 0.583(0.012)& 1.553(0.036) \\
         & -13.727(0.050)& 0.164(0.007)& 1.419(0.125)& 0.445(0.044) \\
         & -10.407(0.005)& 1.851(0.026)& 0.542(0.009)& 1.923(0.041) \\
         & -8.433(0.025)& 0.367(0.009)& 1.335(0.048)& 0.940(0.040) \\
         & -4.633(0.066)& 0.154(0.006)& 2.156(0.139)& 0.637(0.048) \\
         & -2.134(0.000)& 0.095(0.000)& 1.441(0.000)& 0.262(0.000) \\
\hline

\end{tabular}}
\end{table*}

\begin{table*}
\caption[]{HNC Absorption line decomposition products}
{
\begin{tabular}{lcccc}
\hline
Source & v    & $\tau_0$ & FWHM & $\int\tau dv$ \\
       & \kms &          &\kms  & \kms \\
\hline
B0212+735& -0.512(0.035)& 0.341(0.069)& 0.226(0.048)& 0.148(0.044) \\
         & 3.766(0.042)& 0.455(0.024)& 1.159(0.107)& 1.009(0.107) \\
B0355+508& -17.172(0.004)& 0.383(0.006)& 0.577(0.014)& 0.423(0.012) \\
         & -13.713(0.104)& 0.016(0.001)& 1.910(0.270)& 0.058(0.009) \\
         & -11.229(0.150)& 0.020(0.003)& 0.735(0.355)& 0.029(0.015) \\
         & -10.384(0.005)& 0.676(0.009)& 0.498(0.014)& 0.645(0.020) \\
         & -8.564(0.022)& 0.074(0.001)& 1.577(0.060)& 0.223(0.009) \\
         & -4.241(0.063)& 0.024(0.001)& 1.715(0.156)& 0.080(0.008) \\
B0415+379& -2.467(0.004)& 0.367(0.002)& 1.004(0.011)& 0.706(0.009) \\
         & -0.874(0.002)& 1.662(0.006)& 0.743(0.006)& 2.365(0.021) \\
         & 0.127(0.021)& 0.083(0.005)& 0.491(0.064)& 0.078(0.011) \\
B0528+134& 9.382(0.052)& 0.105(0.008)& 0.964(0.146)& 0.193(0.033) \\
B1730-130& 4.890(0.093)& 0.022(0.003)& 0.897(0.275)& 0.038(0.013) \\
         & 6.733(0.305)& 0.007(0.003)& 0.897(0.918)& 0.011(0.013) \\
B2200+420& -1.209(0.061)& 0.217(0.019)& 1.015(0.164)& 0.423(0.078) \\
B2251+158& -10.092(0.190)& 0.018(0.004)& 1.305(0.492)& 0.046(0.019) \\

\hline

\end{tabular}}
\end{table*}

\end{appendix}

\end{document}